\documentclass[a4paper,11pt]{article}
\pdfoutput=1 

\usepackage{jinstpb} 

\usepackage{lineno}

\title{\boldmath Observation of soft X-ray Cherenkov radiation in Be and Si foils}

 \author[1]{S. R. Uglov,\note{ Corresponding author.}}
 \author[]{A.V.Vukolov}
 \affiliation{Tomsk Polytechnic University,\\Lenin Avenue 30, Tomsk 634050, Russia}

\emailAdd{uglov@tpu.ru}

\abstract{Currently, most of the Cherenkov phenomenon researches for the X-ray region are theoretical studies; there are only a few experiments in this area. Here the new experimental results on observation of X-ray Cherenkov radiation generated by 5.7 MeV electrons in the thin Be and Si foils are presented. The Cherenkov effect from Be was observed for the first time. The experimental results are compared with the calculations performed according to the theoretical model of transition radiation taking into account the oxide layer on the target output surface.}


\begin{document}
\maketitle
\flushbottom

\section{Introduction}

It is known that the Cherenkov radiation (CR) is generated by a charged particle moving in a medium with a refraction index n > 1, if the particle velocity $\beta = v/c > 1/n$. The Cherenkov radiation in the optical range is widely known and is used, for example, in charged particle detectors. Also for some substances, the condition n > 1 is met at small spectrum intervals in the region of vacuum ultraviolet ~\cite{aa1,aa2} and soft X-rays near the radiation absorption edges ~\cite{aa3,aa4,aa5,aa6,aa7,aa8}. So the Cherenkov radiation for X-ray (XCR) is possible too. This radiation is considered as an effective method of obtaining radiation in the ultrasoft X-ray range ~\cite{aa8,aa9,a10}, including in the region of the "water window" (E = 280 ÷ 543 eV) ~\cite{aa8,aa9} and “carbon window” (E = 248 ÷ 282 eV), which are relevant in biological research ~\cite{a11}. 

Currently, most of the XCR researches are theoretical studies ~\cite{a12,a13,a14,a15,a16,a17}, and there are only a few experiments in this area. These experiments presented convincing experimental evidence for the observation of the Cherenkov effect in the X-ray range. At the same time, not all experimental results show complete agreement with the theory.

Thus, after the first experimental observation of the Cherenkov effect in the ultra-soft X-ray range near the K-edge of carbon absorption ~\cite{aa3}, Dutch researchers obtained in their study ~\cite{aa6} an experimental confirmation of the Cherenkov effect existing near the absorption edges of elements such as titanium, vanadium, and silicon; at the same time, they failed to detect the Cherenkov effect for Ni and to confirm the Cherenkov effect in carbon ~\cite{aa9}. The carbon-related researches were carried out using various modifications of carbon such as diamond, amorphous carbon, graphite and carbon in organic compounds using a lower electron energy than in ~\cite{aa3}, but sufficient for the appearance of Cherenkov radiation according to the theory.

There are also disagreements between the calculations of the XCR angular density and the experimental observation of the one during the sliding interaction of 75\ MeV electron beam with targets ~\cite{a10}. For example, the experimental Cherenkov angles for carbon and Si turned out to be the same, which does not agree with the theory.

Among the recent experimental studies, the results that was  observed on the inner electron beam  of the special  synchrotron "MIRRORCLE-20" [19] are of particular interest. In this experimental studies, an anomalously high angular density of soft X-ray radiation was observed in the plane perpendicular to the magnetic field of the synchrotron. This effect still has no convincing explanation.

The above-mentioned discrepancies of some experimental results with the theory and the fewness of experimental studies are the motive to new experimental studies of XCR, for example, to take the test on ability the other substances to  generate the  XCR.
So recently we have demonstrated the observation of XCR near the L-edge of Al which was generated by 5.7 MeV electrons ~\cite{a20}.

In this paper, we present new results of observation and research of the XCR for the following two substances: 1) thin Be foil: a next new chemical element in which the observation of XCR is expected; 2) thin Si target foil was used as the control for which experimental confirmation of the XCR existence has been obtained previously ~\cite{aa6}. The obtained experimental results were compared to the calculations performed using the study by Pafomov ~\cite{a21}.

\section{Spectral and angular properties of XCR}
The fundamental possibility of observing the XCR effect in some substance can be determined based on the properties of its dielectric constant $\epsilon(\omega)$
\begin{equation}
\label{eq:x}
\epsilon(\omega)=1+\chi^{\prime}(\omega)+\chi^{\prime\prime}(\omega),
\end{equation}
\noindent
where $\chi^{\prime}(\omega)$ and $\chi^{\prime\prime}(\omega)$ are real and imaginary parts of the dielectric susceptibility. The Cherenkov effect can appear for this substance, if the parameter $\chi^{\prime}(\omega) > 0$. As $\epsilon(\omega) = n(\omega)^2$ ($n(\omega)$ is the refraction index), then, considering that for the X-ray range $n(\omega)\sim1$ and, therefore, $\chi^{\prime}(\omega)\ll 1$ and $\chi^{\prime\prime}(\omega)\ll1$, the condition, that corresponds to the main appearance condition of the Cherenkov radiation for the target material $n(\omega)>1$, shall be $\chi^{\prime}>0$. This condition, for example, for Be, Al, and Si atoms, is met near the corresponding K and L edges of the radiation absorption by matter. 

\begin{figure}[htp]
    \centering
    \includegraphics[width=100mm]{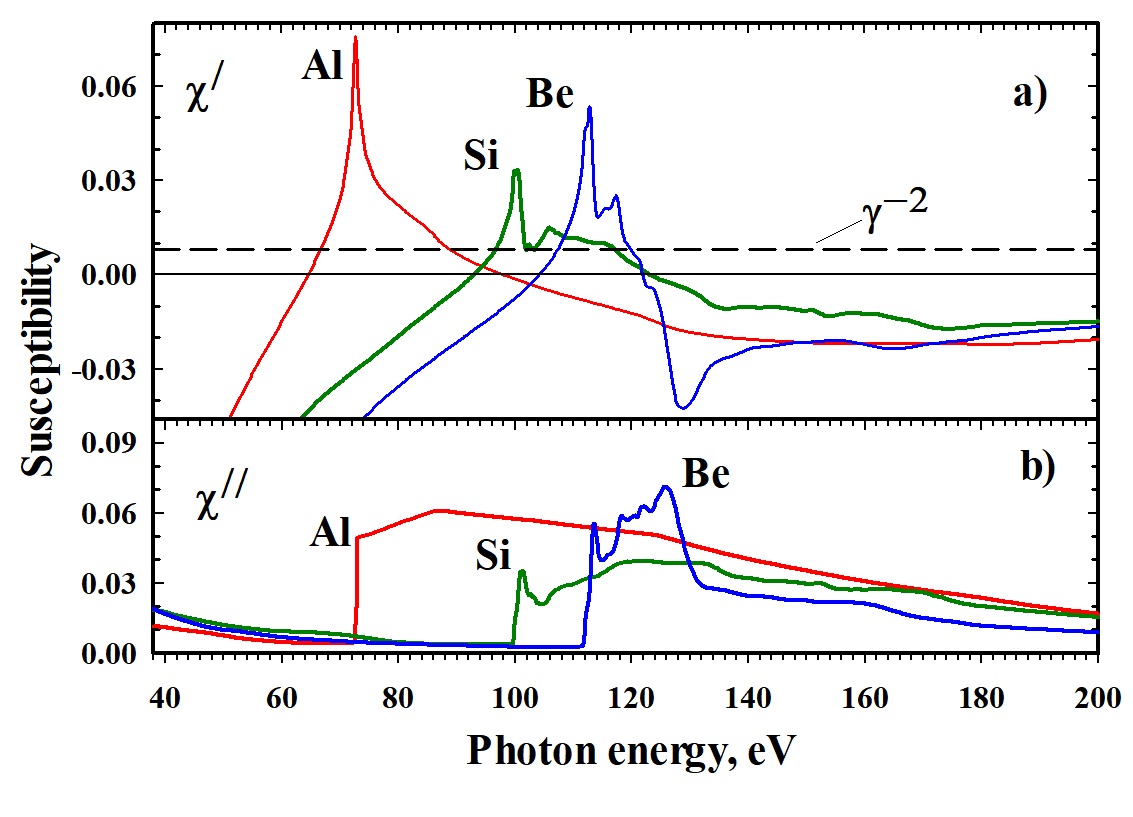}
    \caption{Solid lines are $\chi^{\prime}(\omega)$ (a) and $\chi^{\prime\prime}(\omega)$ (b) for Be, Al, and Si, as calculated using the data of ~\cite{a18}. Dashed line is $\gamma^{-2}$ for electron energy E$_{e}$ = 5.7 MeV.}
    \label{fig:g1}
\end{figure}

Figure~\ref{fig:g1} shows the values of $\chi^{\prime}(\omega)$ and $\chi^{\prime\prime}(\omega)$ for Be, Al, and Si, as calculated using the data of ~\cite{a18}. The second condition for the XCR appearance is the required velocity of a charged particle $v$, that is derived from the condition $\beta = v/c \geq 1/n$ or $\chi^{\prime}(\omega)(\gamma^2-1)\geq 1 $ and if $ \gamma \gg 1$ \ $\chi^{\prime}(\omega)\geq\gamma^{-2}$, where $\gamma$ is the Lorentz factor of the charged particle. The velocity value $v$ with $\gamma^{-2} = \chi^{\prime}(\omega)$ determines the energy threshold of the Cherenkov effect appearance. In figure 1, the dash line marks the value $\gamma^{-2}$ for electrons with energy of 5.7 MeV.

The Cherenkov effect for substances shown in figure~\ref{fig:g1} should appear in the spectral ranges of photon energies between intersection points of the corresponding $\chi^{\prime}(\omega)$ with the dash line $\gamma^{-2}$. However, because of radiation self-absorption by the target substance, due to a sharp increase in absorption near the K or L edge, the actual observation range of the Cherenkov effect on the right side of the spectrum is limited by the energy value corresponding to the absorption edge E$_S$. Therefore, the Cherenkov effect can be actually observed only in the energy interval corresponding to the left slope of the curve $\chi^{\prime}(\omega)$.

In theoretical models of transition radiation, the Cherenkov effect appears automatically when the condition $cos(\theta_{ch}) = 1/\beta n(\omega)$ is satisfied for photon energy E= $\hbar\omega$ and radiation angle $\theta$. When this condition is satisfied, a sharp peak in the spectrum of radiation emitted at the corresponding angle $\theta_{ch}$ appears against the background of the “white” transition radiation spectrum.

The theoretically expected spectrum and distribution of the angular density of the XCR, including the transition radiation generated in the vicinity of the Cherenkov radiation, have been calculated using the formulas from ~\cite{a21} in the semi-infinite target approximation. 

\begin{figure}[htp]
    \centering
    \includegraphics[width=100mm]{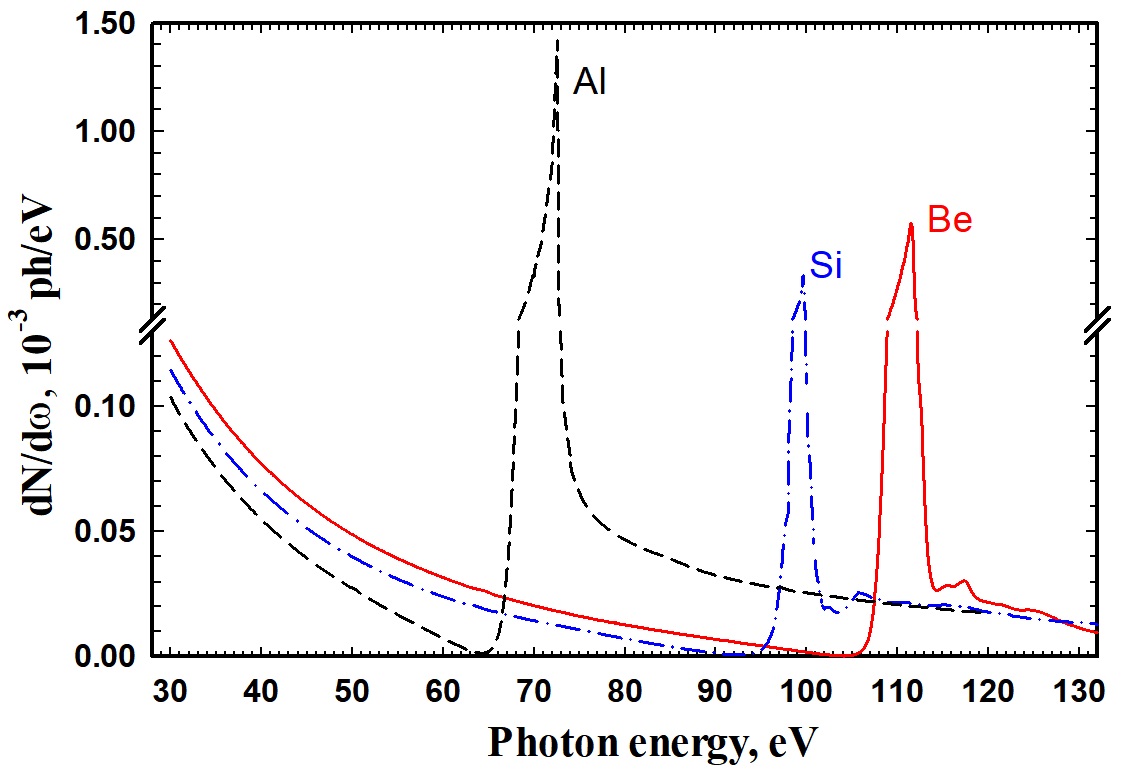}
    \caption{Radiation spectra for Al, Si, and Be targets was calculated using ~\cite{a21} for the transition radiation forward from a semi-infinite target for a (total) electron energy of E$_{e}$ = 5.7 MeV}
    \label{fig:g2}
\end{figure}

Thus, figure~\ref{fig:g2} shows full radiation spectra for Al, Si and Be integrated over the entire radiation cone in the forward direction for electrons with a total energy of E = 5.7 MeV.

\begin{figure}[h!]
    \centering
    \includegraphics[width=8cm]{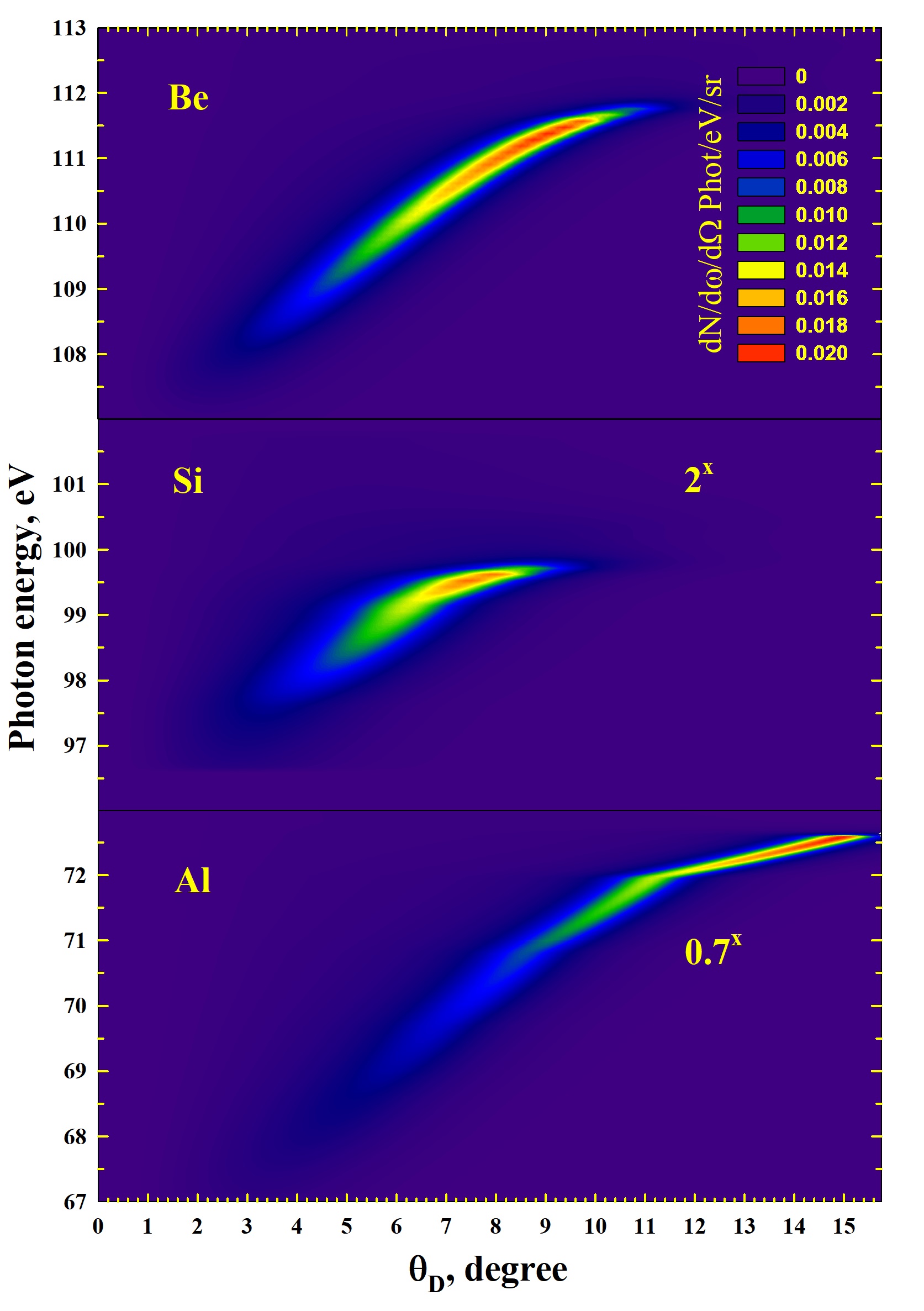}
    \caption{Spectral angular radiation density in the range where the appearance condition is met for the XCR. $ \theta_D $ is the angle between the directions of propagation of the photon and of the electron.}
    \label{fig:g3}
\end{figure}

Figure~\ref{fig:g3} shows the calculated spectral-angular radiation densities for Al, Si and Be  for the range when the conditions $\beta n(\omega) \geq1$ are satisfied.  The 2D patterns illustrate the general properties of XCR such as the energy range and the line width of the spectrum of the XCR photons, the angular width of the XCR cone and the structure of the XCR intensity distribution within the cone. The values of $dN/d\omega/d\Omega$ for Be, Si and Al are presented using scale factors 1, 2 and 0.7, respectively. The table~\ref{tab:i} shows the photon energy E$_{max}$, the emission angle $\theta_{max}$ and the yield corresponding to the maximum spectral-angular density. The  E$_{L}$ and E$_{R}$ in the table~\ref{tab:i}, indicate the boundaries of the spectral interval meeting the XCR appearance condition. 

\begin{figure}[htp]
    \centering
    \includegraphics[width=48mm]{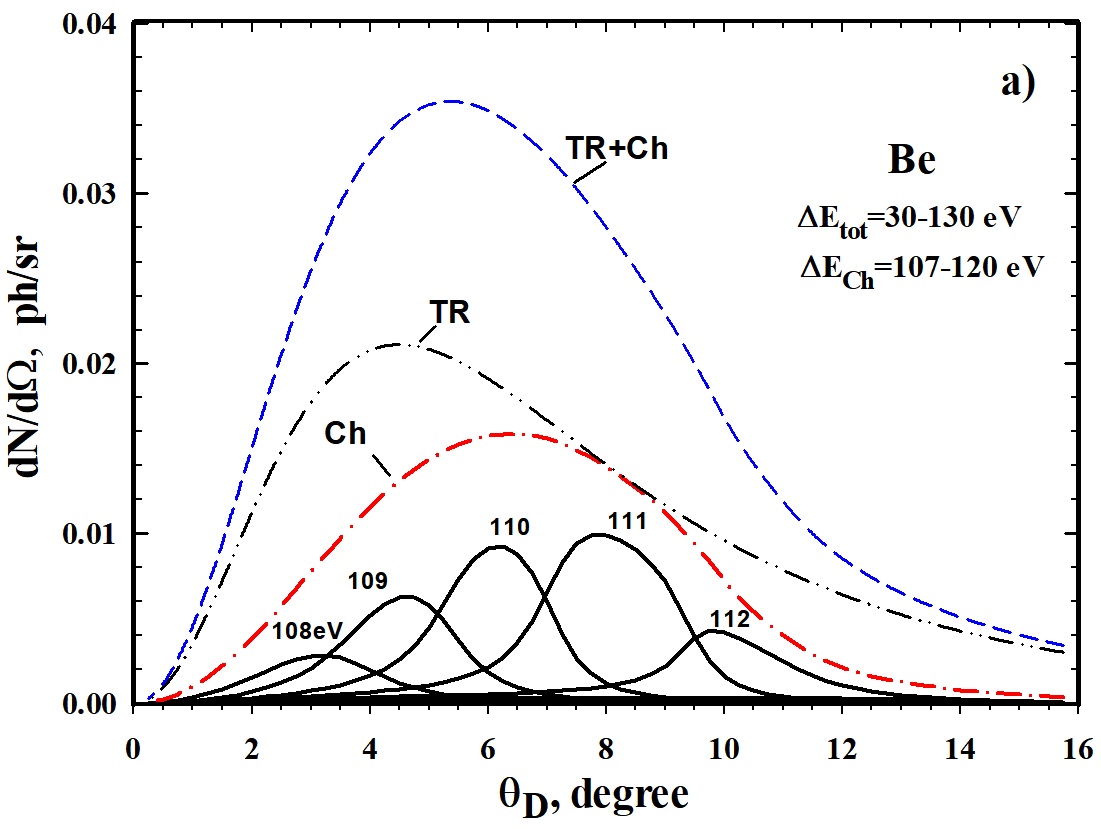}
    \includegraphics[width=48mm]{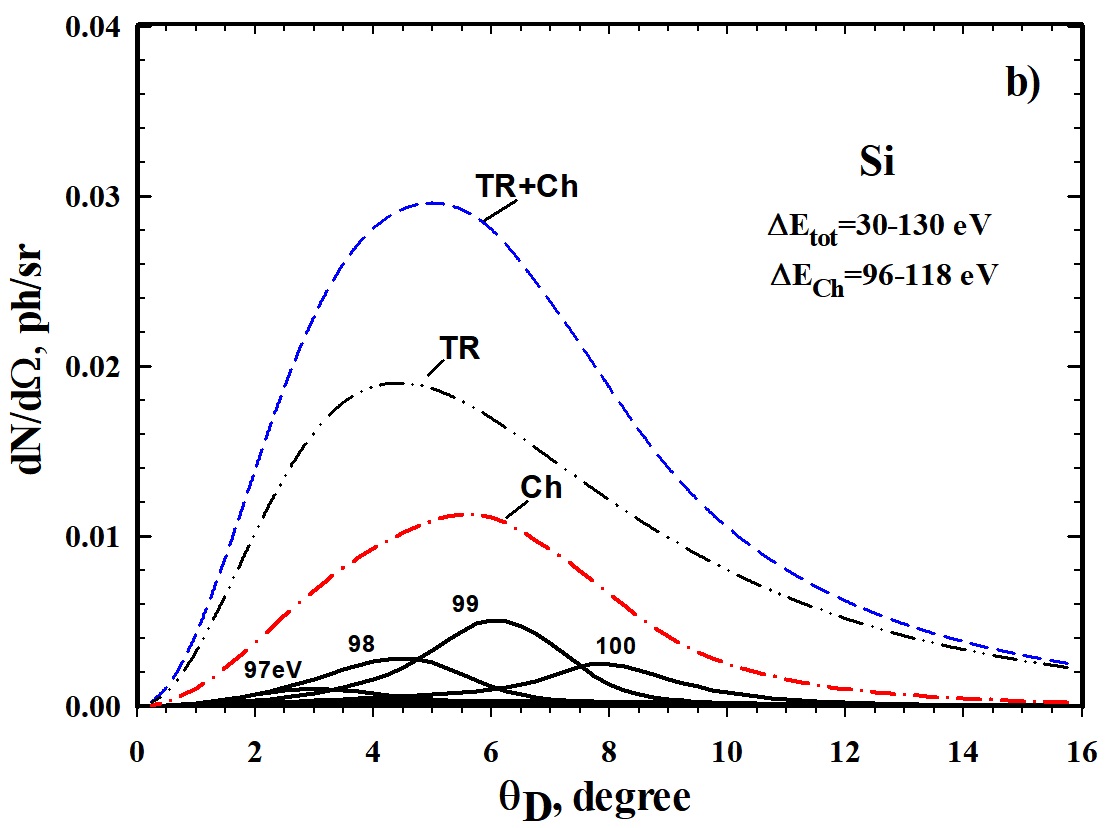}
    \includegraphics[width=48mm]{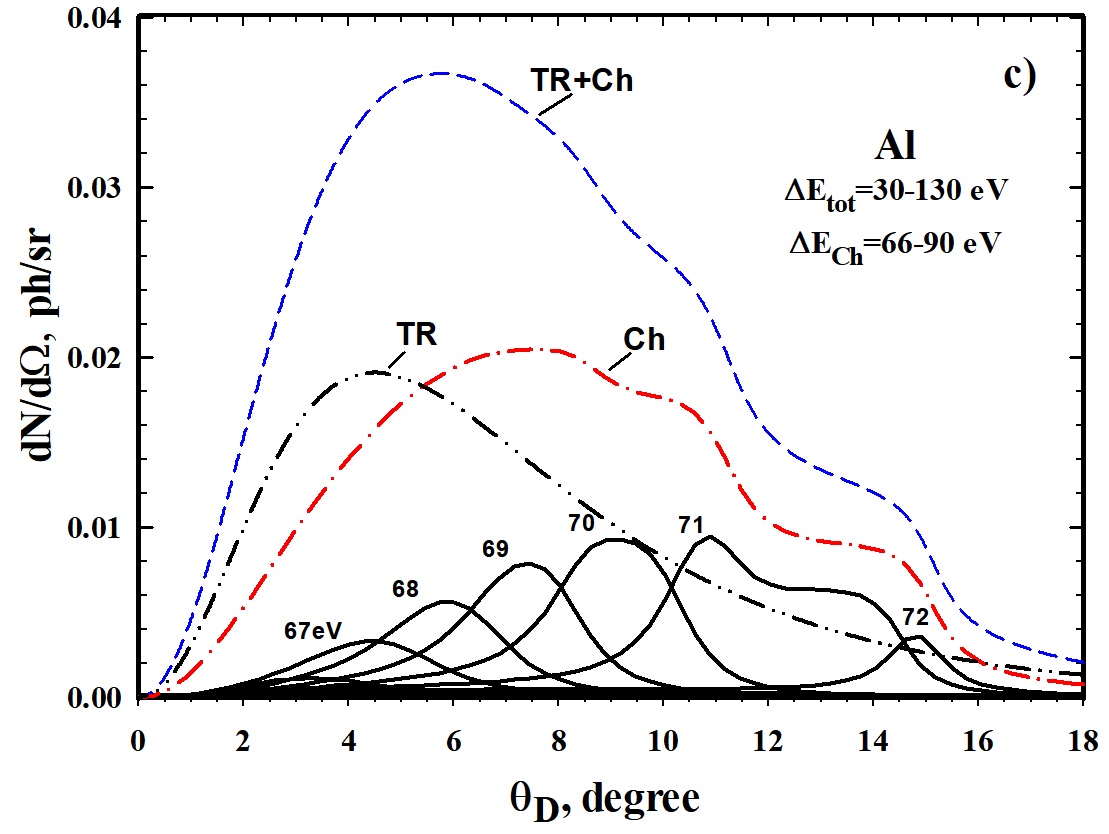}
    \caption{Components of the angular distribution of the Cherenkov (Ch) and transition (TR) radiations for Be (a), Si (b), and Al (c) depending on the photon energy. Solid lines are calculated for $\bigtriangleup$ E = 1  eV; the dash-dot line is the contribution of Ch; the dash-double-dott line is the contribution of TR outside the Cherenkov interval; the dotted line is TR+Ch within the range of E = 30$\div130$\ eV.}
    \label{fig:g4}
\end{figure}

The calculation results in figure~\ref{fig:g4} show the angular radiation distributions in the spectral range of $\bigtriangleup E=30\div130$ eV as calculated for Al, Be, and Si, both within the photon energy range meeting the Cherenkov condition (dash-dot lines) and being out of this range (dash-double-dot lines), when only the transition radiation is generated.

The solid lines in figure~\ref{fig:g1} correspond to the angular distribution components as calculated for a series of photon energies within the spectral range with the width of $\bigtriangleup$ E = 1 eV within the Cherenkov range: $\bigtriangleup$ E$_{ch}$= E$_{L}\div$ E$_{R}$. Note that in the provided calculations the radiation generated within the Cherenkov energy range is not a pure Cherenkov radiation, but it has an addition from the transition radiation generated at the target’s output surface.

The calculations in figure~\ref{fig:g3} and ~\ref{fig:g4} show that the total angular distribution of the transition and Cherenkov radiations has no pronounced narrow “Cherenkov cone” that is characteristic of the Cherenkov radiation within the optical range ~\cite{a22}. The value of $\theta_{max}$ corresponding to the maximum of the spectral-angular density of XCR is different from the value of $\Theta_{max}$ – the maximum radiation of the angular density $dN/d\Omega$ (see  table~\ref{tab:i} columns 6 and 9). The angular distributions of the transition radiation generated over the entire spectral interval under consideration and that of XCR virtually coincide both in intensity and in the emission direction (see table~\ref{tab:i}). Therefore, the calculations of spectra and angular distributions show that it is difficult to experimentally prove the Cherenkov effect in the X-Ray range based on the analysis of the angular distribution shape. A more convincing result can be obtained by spectral research, by finding an intense spectral line in the spectrum of radiation emitted in the direction of the expected Cherenkov radiation cone, with the energy virtually coinciding with the energy of the corresponding absorption edge. 
\begin{center}
\begin{table}[h!]
\footnotesize
\centering
\caption{\label{tab:i} Spectral and angular properties of XCR and TR}
\smallskip
\begin{tabular} [b] {|c|c|c|c|c|c|c|c|c|c|c|} \hline
\multicolumn{3}{|p{1.18in}|}{XCR spectral range with E$_e$ = 5.7 (MeV)} &  \multicolumn{6}{c|}{\small{Cherenkov radiation}} &
\multicolumn{2}{p{1.19in}|}{{Transition radiation, $ \bigtriangleup$E=30$\div 130$ (eV) }}\\
\hline
1 & 2 & 3 & 4 & 5 & 6 & 7 & 8 & 9 & 1 0 & 11\\
\hline
\multicolumn{1}{|c|}{}&  E$_L$ & E$_R$ & E$_S$ & dN/d$\omega$/d$\Omega$ & $\theta_{max}$ & E$_{max}$ & dN/d$\Omega$ & $\Theta_{max}$ & dN/d$\Omega$ &  $\theta_{max}$\\
\multicolumn{1}{|c|}{}& (eV) &(eV)&(eV)& (ph/sr/eV)&(degr)&(eV)&(ph/sr)&(deg) & (ph/sr) & (deg)\\
\hline
Al & 66.7	&89.5	&72.6&	0.0287&	14.63&	72.52	&0.0205&	7.6	&0.0191&	4.44\\
\hline
Si	&96.6&	117.5&	99.8&	0.0091&	7.5&	99.52& 0.0113	&5.75	&0.0190	&4.38\\
\hline
Be	&107.4	&120.1&	111.5&	0.0201&	8.75&	111.29&	0.0159&	6.25&	0.0211&	4.5\\
\hline
\end{tabular}
\end{table}
\end{center}

\section{Method and setup}
\label{sec:intro}
\begin{figure}[htp]
    \centering
    \includegraphics[width=12cm]{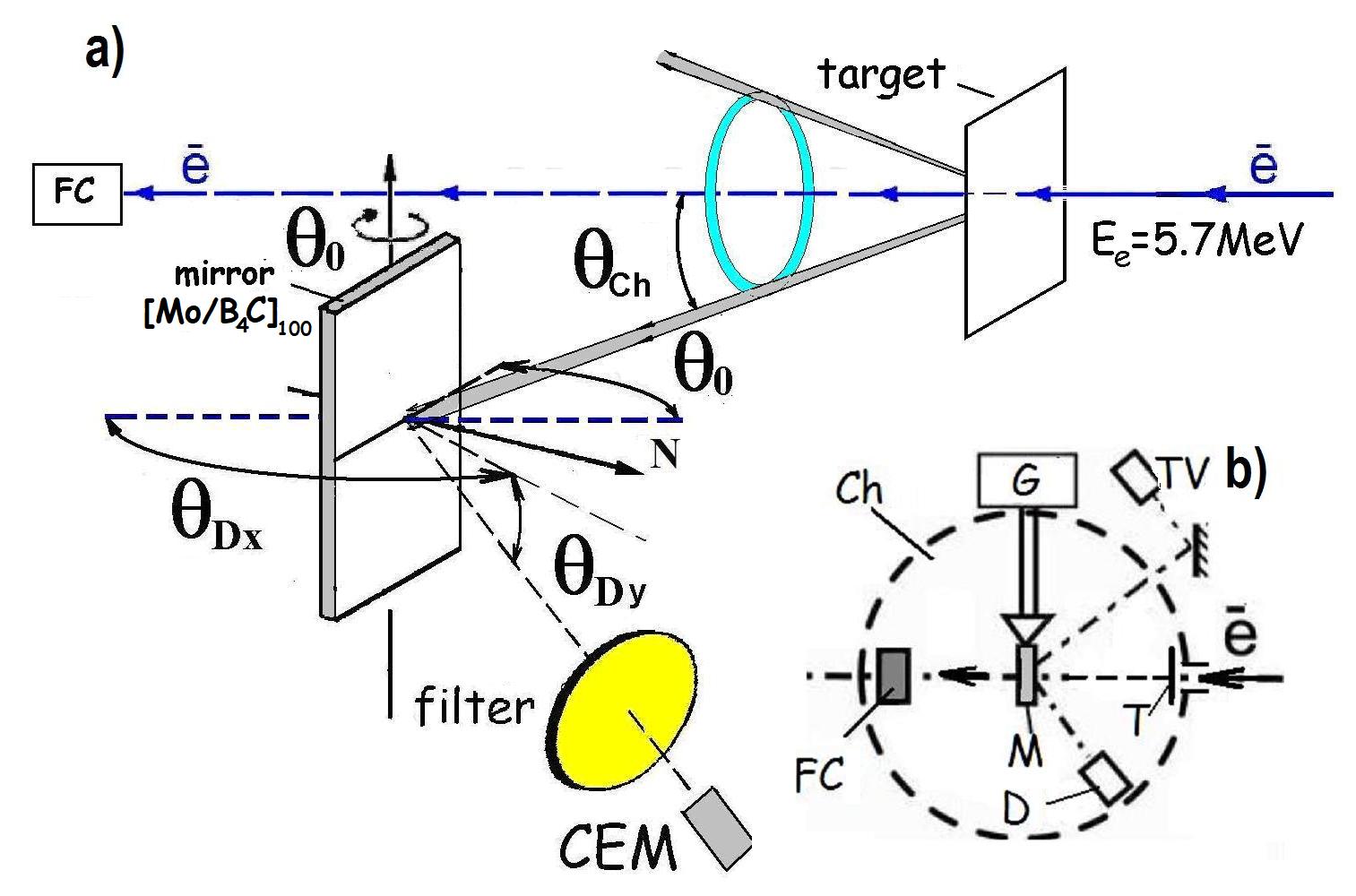}
    \caption{a) Scheme for XCR investigation using a multilayer mirror; b) top view of the vacuum chamber - Ch; T is a target, D is a channel-electron multiplier (CEM), M is a multilayer mirror, G is a goniometer, FC is a Faraday cup.}
    \label{fig:g5}
\end{figure}

In the research of the radiation spectral composition, we used the Bragg scattering of radiation by a multilayer mirror. Figure~\ref{fig:g5}a shows the XCR observation setup using the Bragg reflection of a radiation cone fragment near the angle $\theta_{Ch}$ using a multilayer mirror [Mo/B$_4$C]$^{100}$ with a period d = 7.56 nm to select the radiation by photon energy.

\begin{figure}[htp]
    \centering
    \includegraphics[width=75mm]{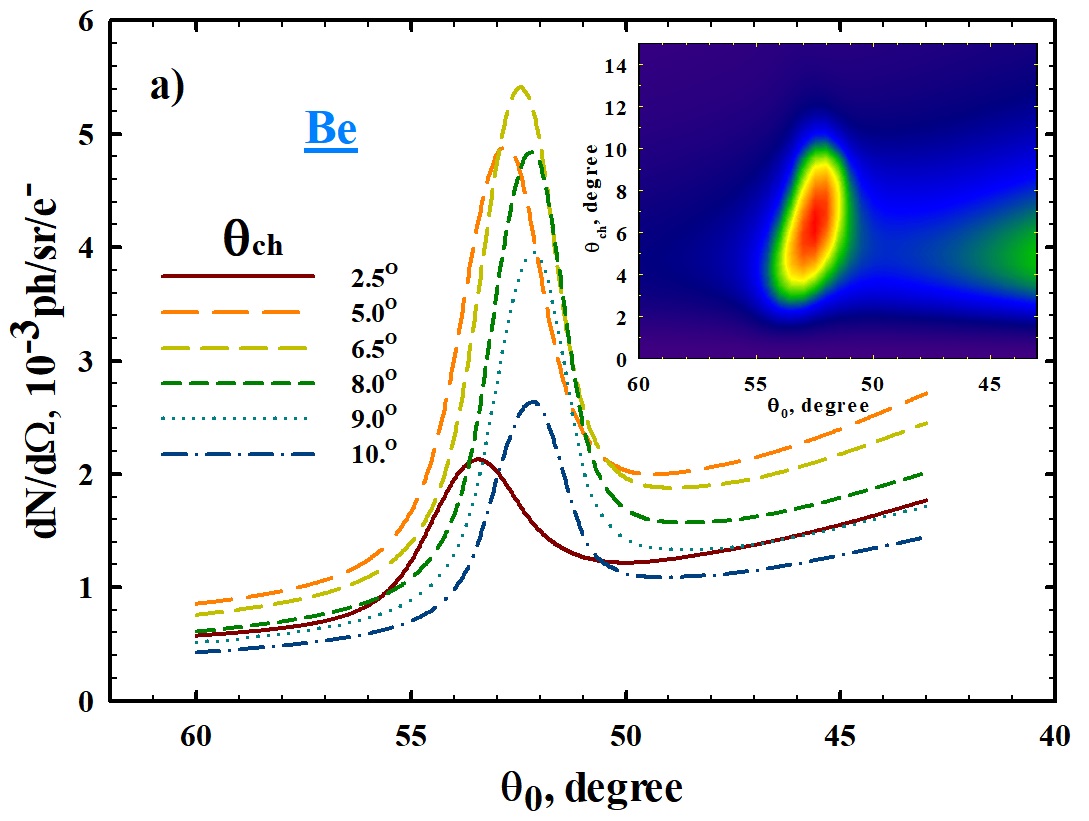}
    \includegraphics[width=75mm]{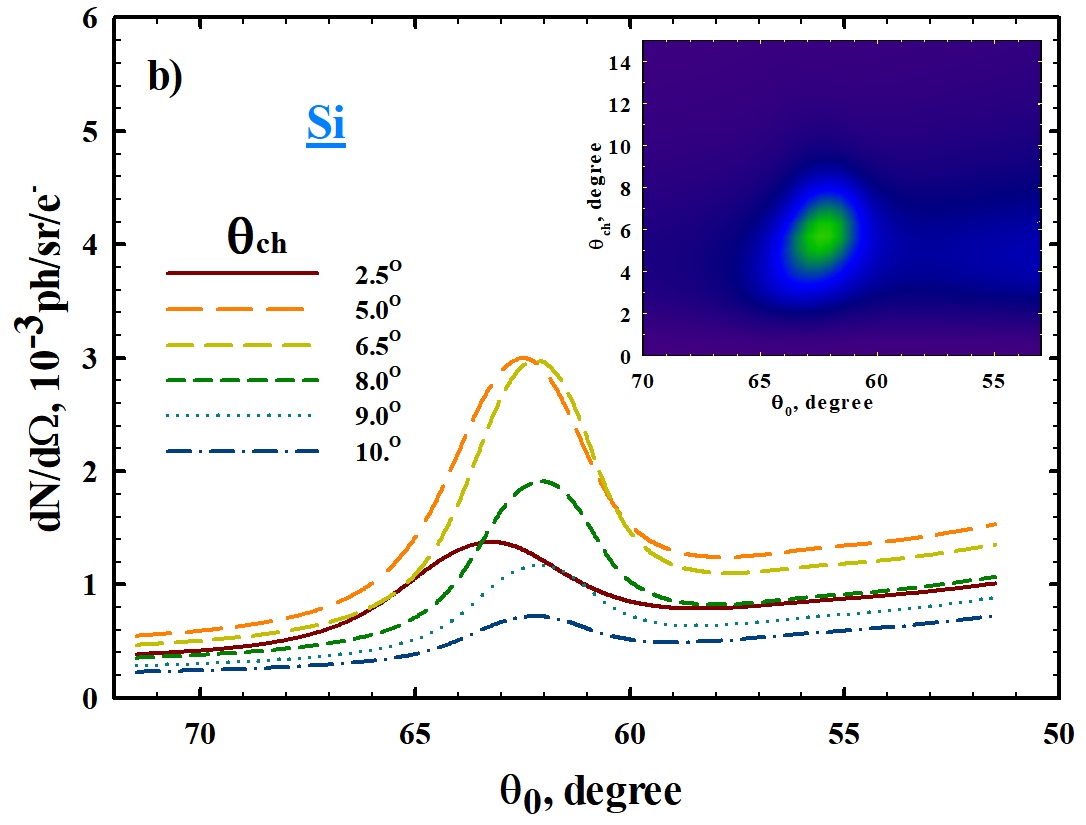}
    \caption{Intensity of radiation reflected by a multilayer mirror [Mo/B$_4$C]$^{100}$ depending on $\theta_0$ : a) for Be and b) for Si targets.}
    \label{fig:g6}
\end{figure}

The results of calculations of the radiation intensity reflected by a multilayer mirror, depending on the angle $\theta_0$, are shown in figure~\ref{fig:g6}. The calculations have been performed for radiation emitted along of the radiation plane that coinciding with the mirror rotation axis $\theta_0$. The calculations have taken into account the reflection index of the multilayer mirror as calculated using ~\cite{a23,a24,a25} by the recurrence relation method for the spectral-angular density of the transition and Cherenkov radiations calculated according to ~\cite{a21} for the spectral interval $\bigtriangleup$ E = 30$ \div 400$ eV.

The maximum values on the curves correspond to the XCR photon energy. For a target that generates only transition radiation, for example, for a Mylar film~\cite{a20}, no maximum should be observed.
The experiment was carried out on a 5.7 MeV electron beam extracted from the M-5 microtron of the Tomsk Polytechnic University ~\cite{a26}. The top view of the scattering chamber is shown in figure~\ref{fig:g5}b. The accelerator operating frequency was 50 Hz; the duration of the macro impulse of the extracted electron beam (MIB) was 0.4 $\mu$s. In the scattering chamber, the beam had a circle shape with $\oslash = 2 \div 3$ mm. The average number of accelerated electrons, as measured using a Faraday cup installed behind the target, was $2.5\times 10^7$ per MIB. 

The targets were located at the entrance to the scattering chamber. In both cases, the target thickness was much greater than the length of the radiation absorption by the target at the Cherenkov radiation energy. The beryllium target had the thickness of t$_{Be}$ = 26 $\mu$m; the silicon target, t$_{Si}$=~\ 4$\ \mu$m. The transverse dimensions of the targets were 20$\times$20\ mm and 6$\times$8\ mm, respectively.

In the center of the scattering chamber, at the distance of t$_m$ = 200 mm from the target, there was a multilayer mirror installed.
The mirror working size was formed using a special mask made of Mylar with rectangular vertical diaphragm 10×35 mm$^2$ in size, installed directly on the mirror body.
The mirror [Mo/B$_4$C]$^{100}$ consisted of 100 pairs of Mo and B$_4$C layers on a silicon substrate.
The mirror period was d = 7.56 nm; the ratio of the B$_4$C layer thickness to the mirror period was $\Gamma$~=~0.5. 
The composition of the transition layers was estimated as consisting of a mixture of 5Mo+B$_4$C with a thickness of $\bigtriangleup$t = 0.9 $\div$ 0.95 nm.
On average, the mirror reflection index, as calculated using the IMD-5 ~\cite{a27} package in the specular scattering approximation, was 30$\div$35\ \% in 
the photon energy interval $\bigtriangleup$ E = 80$\div$120 eV.

The rotation axis of the multilayer mirror crossed the electron beam axis at a right angle. The mirror upper edge was 20 mm below the electron beam axis. 

The radiation reflected by the multilayer system was recorded by a spiral channel electron multiplier (CEM) with a funnel (make: SEM-6). The CEM operated in the counting mode, keeping the linearity at loading up to 2 pulses per MIB. The CEM loading was monitored per 100 MIB of the accelerator and, generally, was 10 to 20 pulses per 100 MIB. 

The detector recording efficiency, as averaged over an entrance window 9 mm in diameter, was 6.9 \% at the photon energy of 96.7 eV and 6.5 \% at the photon energy of 77.1 eV. The lower threshold of the detector is about $E_{low}$ = 12 eV ~\cite{a28,a29,a30}.

The detector was at the distance of L$_D$ = 144 mm from the mirror. Between the detector and the mirror, we installed a block of permanent magnets to protect the detector from scattered electrons and a nitrocellulose (NC) plastic filter with a thickness of about 0.2 $\mu$m to suppress the soft component $\hbar\omega$<50 eV of the transition radiation.

To eliminate the effect of scattered magnetic fields on the electron beam, the permanent rare-earth magnets were installed inside a steel case (SC). The SC was placed at 50 mm from target, had length along the radiation registration direction of 80~\ mm and had the through hole of $\oslash$ = 15 mm, that served as a preliminary collimator. The strength of magnetic-field within SC was about H=0.1 T. The radiation from the mirror was additionally collimated by two slits with widths of 7 mm and 8 mm that placed in the center and at the out hole of the SC, respectively.

The axis of the radiation registration channel between the mirror and the detector is formed by the diaphragms in the direction of mirror reflection of the fragment of the cone radiation corresponding to $\theta_{Ch}$ = 6.5$^o$. The angular sizes and solid angle of the radiation cone fragment registered by the detector were $\bigtriangleup\theta_{Ch}^x= \pm 0.67^o$, $\bigtriangleup\theta_{Ch}^y= \pm 0.74^o$ and $\Omega = 5.2 \times 10^{-4}$sr, respectively. The detector rotation axis coincided with the mirror rotation axis.

\begin{figure}[htp]
    \centering
    \includegraphics[width=80mm]{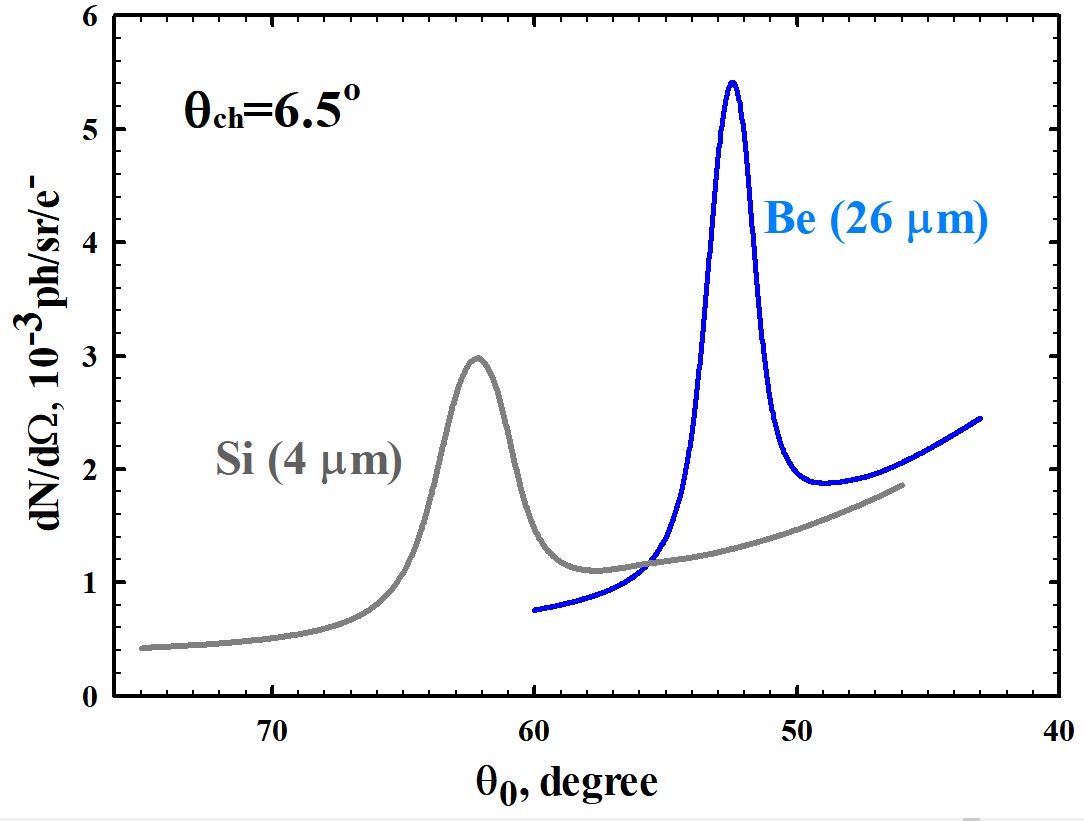}
    \caption{Intensity of the reflected radiation as a function of the angle $\theta_0$ of the [Mo/B$_4$C]$^{100}$ mirror for Si  and Be targets; $\theta_{Dy} = 6.5^o$.}
    \label{fig:g7}
\end{figure}

Figure~\ref{fig:g7} shows the expected intensity of the radiation reflected by the multilayer mirror [Mo/B$_4$C]$^{100}$ for Be and Si targets as a function of the inclination angle $\theta_0$ at $\theta_{Ch} = 6.5^o$.

\section{Results and discussion}

The intensity of the radiation reflected by the multilayer mirror in the Bragg direction was found for each $\theta_0$ from the results of the initial measurement of the radiation intensity as a depending on the observation angle $\theta_{Dx}$. The maximum values of the measured angular distributions after subtracting the background were used for plotting the intensity of the reflected radiation vs $\theta_0$. 
\begin{figure}[htp]
    \centering
    \includegraphics[width=80mm]{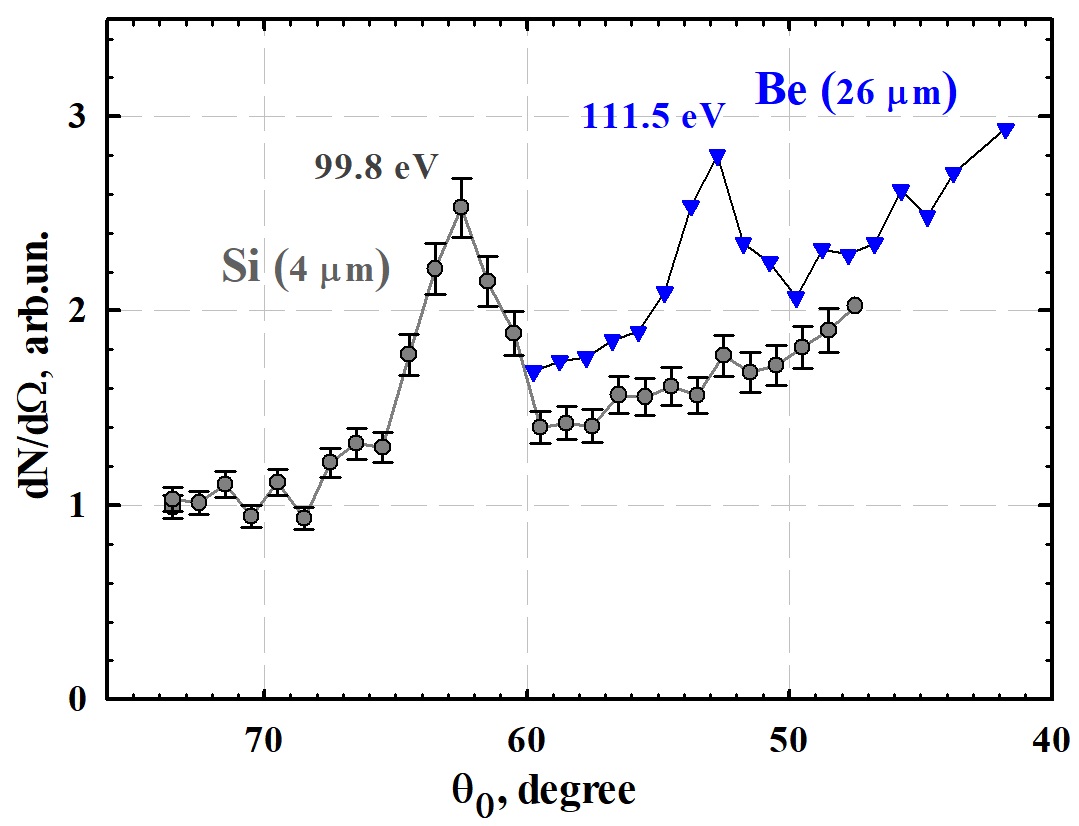}
    \caption{Intensity of the reflected radiation as a function of the angle $\theta_0$ of the  [Mo/B$_4$C]$^{100}$ mirror for Si (grey circles) and Be (blue triangles) targets; $\theta_{Dy} = 6.5^o$.}
    \label{fig:g8}
\end{figure}
Figure~\ref{fig:g8} shows the maximum intensity values of the radiation reflected by the mirror as obtained for Si and Be targets. As one can see, the general view and position of the maxima of the experimental curves are similar to the curves in figure~\ref{fig:g7}. The maximum values on the curves at angles of $\theta_0 = 62^o$ and $\theta_0 = 52^o$ correspond to the energy E$_{Si}$ = 99.8 eV (the bottom curve) and E$_{Be}$ = 111.5 eV (the upper curve). The values E$_{Si}$ and E$_{Be}$ are quite consistent with the L and K absorption edges of silicon and beryllium, respectively. Therefore, we can affirm that we experimentally observed the X-ray Cherenkov radiation for the first time in a target made of Be and confirmed the X-ray Cherenkov radiation in Si.

 However, when comparing the  experimental results with the calculations in detail, we see a discrepancy  between the ratio of the intensities  for Si and Be and also there is some background pedestal in the experiment.

We consider that one of the main reasons for such difference between theory and experiment is the presence of the oxide layer or another contamination on the target surface. To take into account for the effect of an additional layer, we carried out calculations using Pafomov’s expressions for a target consisting of three substances. It is semi-infinite metal/oxide/vacuum in our case.  The original expression valid for an arbitrarily moving charged particle has a complex form,  and one may find them in ~\cite{a21}. The expression for the spectral density of photons per unit solid angle that is valid for the special case of a normal incidence of the charged particle are presented in Appendix {\bf{A}}. In this case, the generated radiation is linearly polarized in the propagation plane, i.e. the electric vector lies in the plane formed by the velocity of the charged particle and the wave vector of the emitted photon. 

\begin{figure}[htp]
    \centering
    \includegraphics[width=80mm]{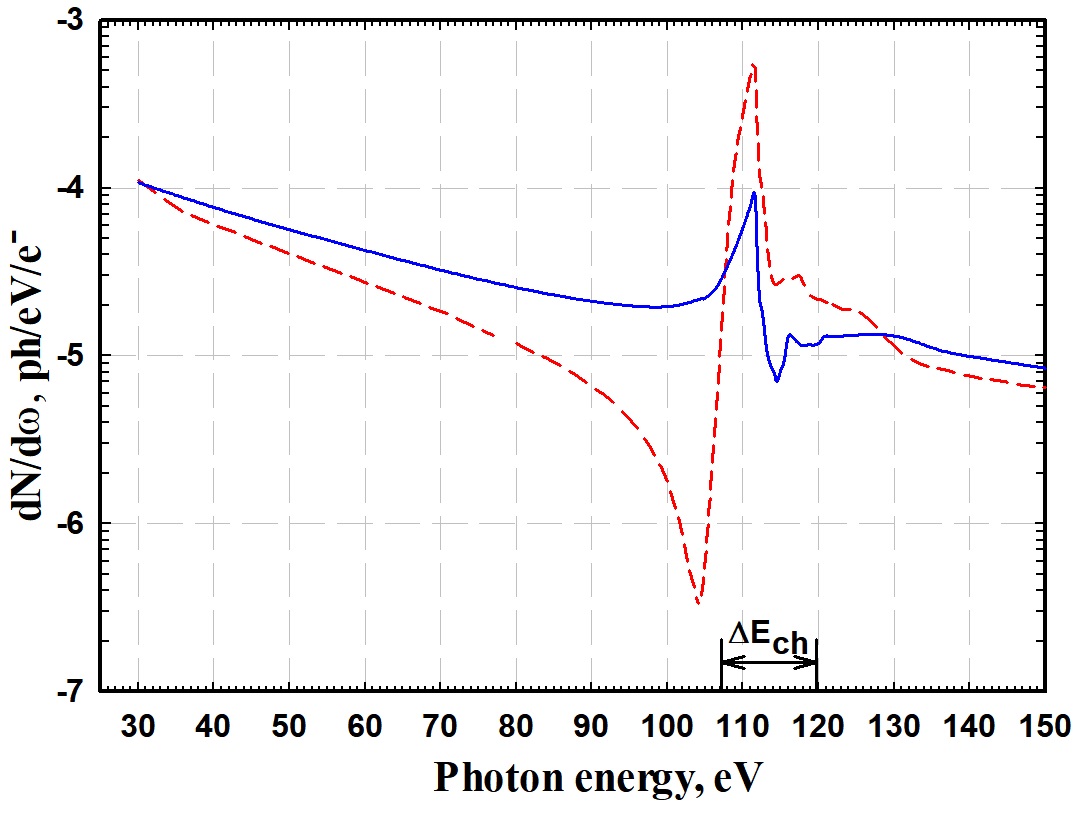}
    \caption{Radiation spectra for: dash line - Be target; solid line - Be targets with a 180~\ nm thick BeO layer.}
    \label{fig:g9}
\end{figure}

The calculations were carried out using a homogeneous layer of BeO or SiO$_2$ oxides as an additional layer on the target surface. The results of calculating the effect of the oxide film on the total radiation spectrum using beryllium as an example are shown in figure~\ref{fig:g9}. Calculations have been performed for a 180~\ nm thick homogeneous oxide layer.

In calculations for BeO, we took into account the chemical shift which leads to a shift of the beryllium absorption edge in combination with oxygen by 2.8 eV towards a higher energy ~\cite{a31,a32,a33}. The chemical shift of the absorption edge in SiO$_2$ is 4.5 eV. As can be seen in figure~\ref{fig:g9}, the effect of the oxide layer on the emission spectrum manifests itself in the suppression of Cherenkov radiation and in the increase of the transition radiation.

Note that in the case of oblique interaction of the electron beam with the target, the suppression of the Cherenkov radiation yield should be expected due to an increase in the effective oxide thickness.

Concerning the background pedestal in the experiment the simple estimations showed that the pedestal radiation intensity is much higher than the contributions from bremsstrahlung and characteristic radiations. The most plausible explanation for the pedestal is the presence of pinholes in the NC filter.
In this case, the high intensity of the pedestal is due to the contribution from  the ordinary reflection of the wide band of high intensity of the soft part (E< 60 eV) of the transition radiation that passed through the pinholes in the NC filter. 

\begin{figure}[htp]
    \centering
    \includegraphics[width=80mm]{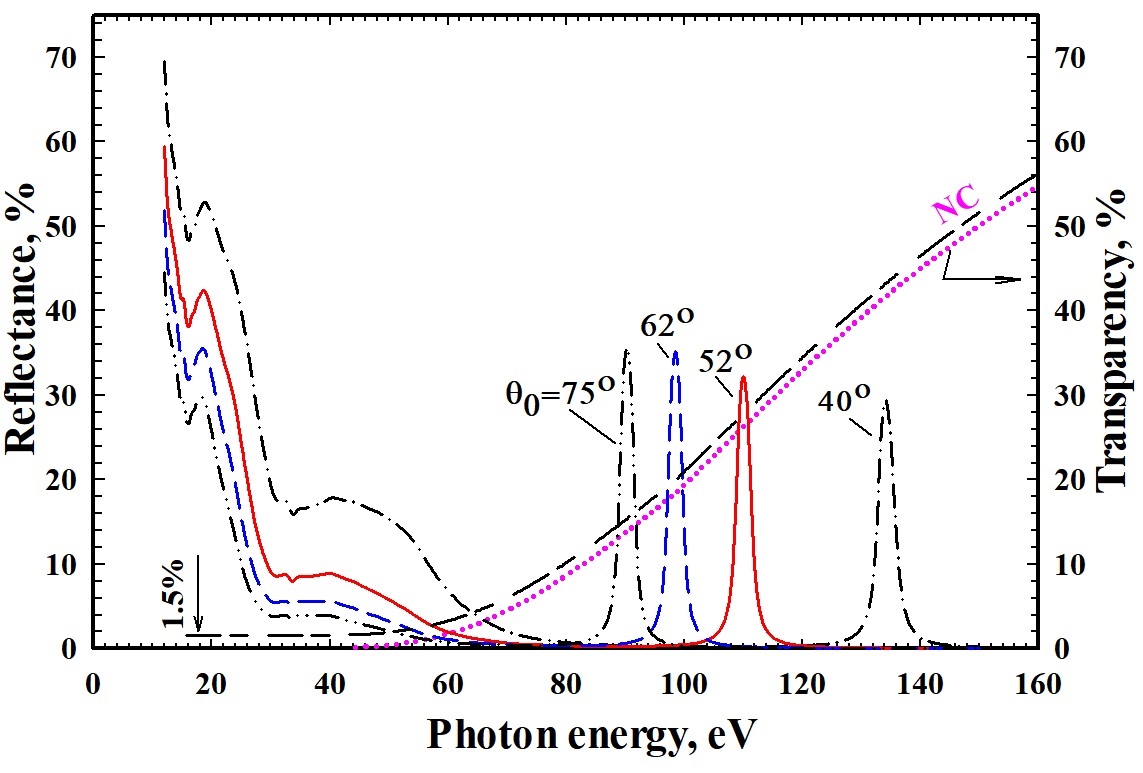}
    \caption{The solid and dashed lines are the reflectance spectra of the   [Mo/B$_4$C]$^{100}$ mirror for several Bragg angle ($\theta_0 =75^o, 62^o, 52^o$ \ and $40^o )$; dotted line is the transparency of NC filter; short-long dashed line is the  transparency of NC with pinholes.}
    \label{fig:6add}
\end{figure}

In order to take into account the contribution due to the pinholes, the emission spectra for Be and Si were calculated starting with the photon energy E = 12 eV, that we have taken as  the lower threshold of the detector efficiency. Since for the optical constants in ~\cite{a18} the lower bound of the photon energy is E = 30 eV, calculations for photons with energy E = 12-30 eV were carried out using the database ~\cite{a34}.
Figure ~\ref{fig:6add} shows the reflectances of the mirror and the transparency of the  NC filter as a function of the   photon energy.

\begin{figure}[htp]
    \centering
    \includegraphics[width=80mm]{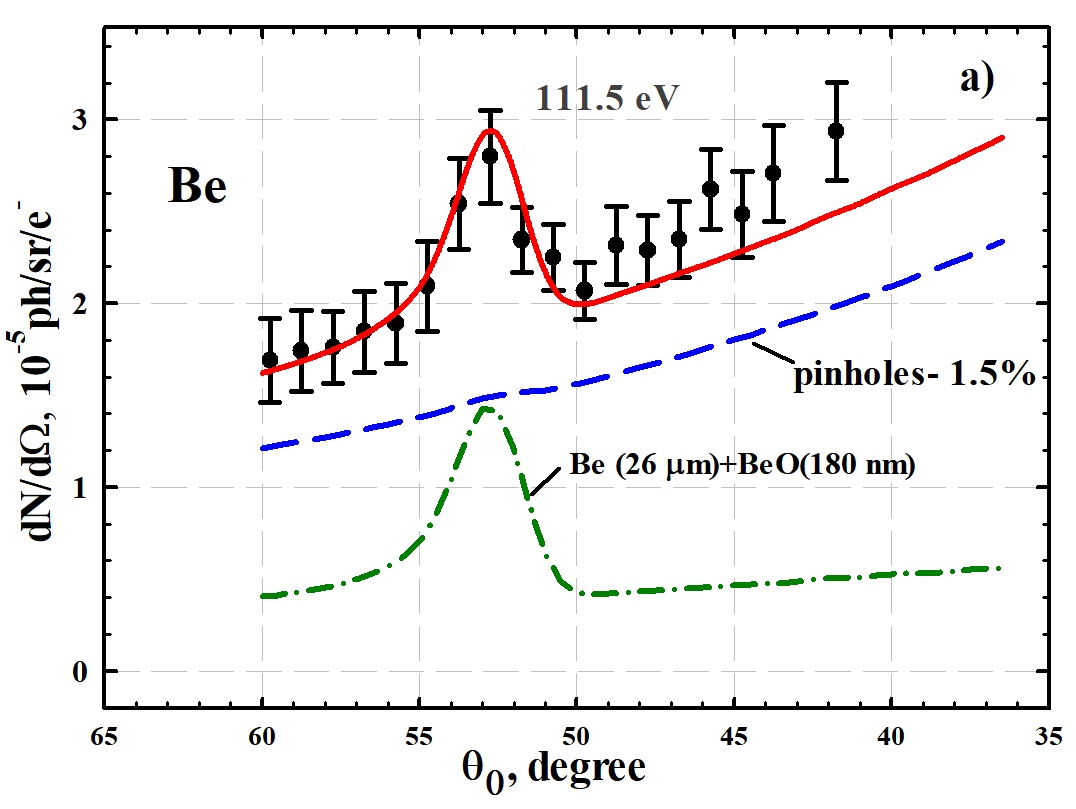}
    \includegraphics[width=80mm]{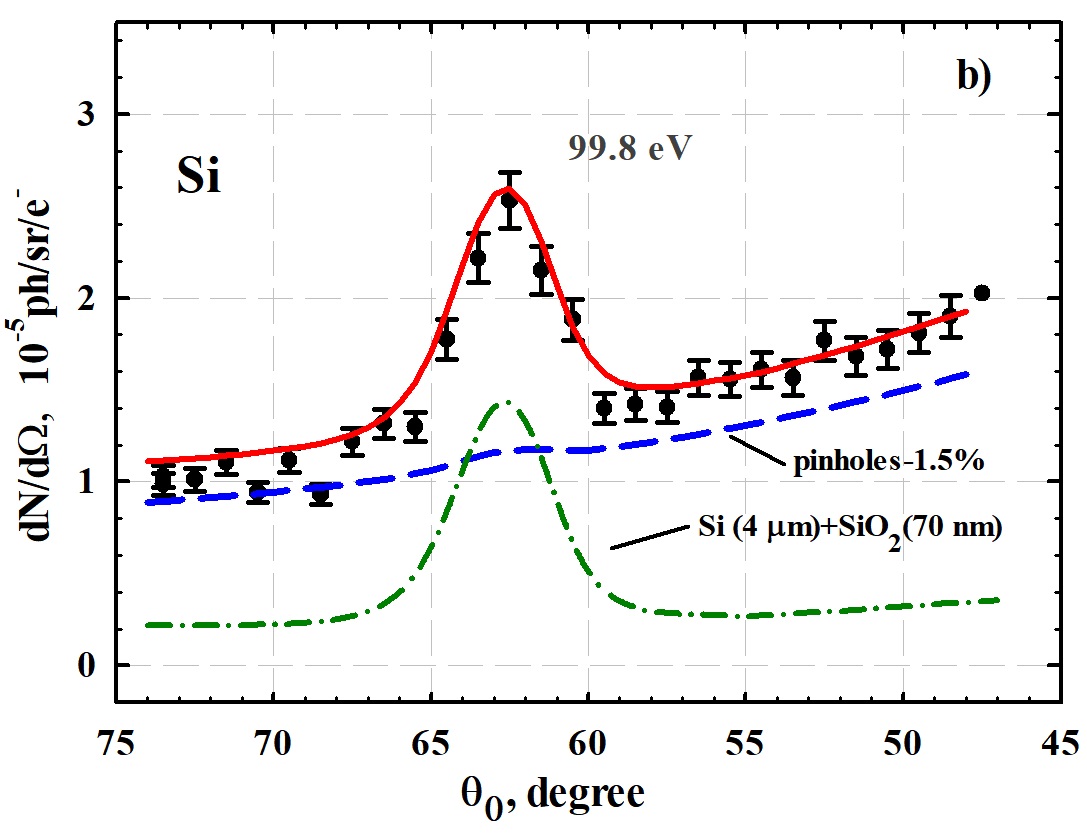}
    \caption{Comparison of calculation and experimental results of the scanning of the radiation spectrum from Be and Si targets by a multilayer mirror.}
    \label{fig:g10}
\end{figure}

Figure ~\ref{fig:g10} shows the experimental results as compared with the calculation results obtained for a three-component medium according to ~\cite{a21} taking into account the  detector efficiency, the NC filter transparency, the efficiency of radiation reflection by the mirror and the thickness of the oxide film. The good agreement between the calculation results and the experiment was obtained by selection of  the thickness of the oxide layer and the fraction of the radiation transmitted through the pinholes. The calculation results presented  by the solid lines were obtained at an oxide layer thickness of 180 nm and 70 nm for Be and Si respectively, taking into account that the total pinhole aperture in the NC filter was 1.5\% of the total detector aperture. The beam divergence was taken into account as normal distribution with $\sigma^2_e=10^{-4}$. The dashed curve in the figure~\ref{fig:g10} shows the contribution associated with radiation transmitted through pinholes. The transparency of the NC filter with pinholes was shown by short-long dashed line in figure ~\ref{fig:6add}. 
.

To verify the presence of an oxide layer on the target surface, the additional study of the elemental composition of the Be target surface using the radio-frequency glow discharge method (RF-GD-OES) was carried out by “GD- PROFILER-2” which showed the qualitatively confirmed presence of a BeO layer with an average thickness of 0.1 to 0.5 $\mu$m on the target surface. The thickness of the oxide layer for a silicon target was not tested.

\section{Conclusion}
As we showed above, the calculations that include the effect of the oxide layer and pinholes, demonstrate an improvement in the agreement between the calculations and the experiment.  However, it should be noted that these calculations were performed using some simplifications and idealization of the parameters of the multilayer mirror and targets. So, for example, the mirror reflectivity was calculated for an ideal periodic structure, the model of the target did not take into account the change of the oxide concentration with increasing distance from the target surface. Besides it's necessary to keep in mind the well-known fact that at low energies of X-ray radiation tabular data for Re$(\chi)$ and Im$(\chi)$ have high uncertainties ~\cite{a34,a35}.

   	Yet, despite the inaccuracies of the models of the multilayer mirror and the surface layer of the target, we can conclude that the results of our studies confirm the existence of the Cherenkov effect for Be and Si. And the Cherenkov effect observation in Be has been presented for the first time. The measured radiation intensities for Be and Si are comparable to the theoretically expected values. The fact of observing the Cherenkov radiation in Be may promote the development of a high-intensity, high-monochrome radiation source with the energy of emitted photons E$\sim$ 111 eV. Following ~\cite{a13}, the spectral-angular radiation density of such a source can be increased several times by using the sliding interaction of the electron beam with the target. Besides, the threshold nature of the XCR can be used for the development of threshold counters for the separation of the charged particles. 	These detectors are more promising for use with multiply charged ions since the XCR yield is proportional to the square of the particle charge.

\appendix
\section {Normal incident for three substances }

The thicknesses of the Si and Be foils used in the experiment are several times greater than the absorption length for the spectral range under study. Therefore, to interpret the results, we used the results of the theoretical paper ~\cite{a21} for radiation generated in the  target that is composed of three matters and two flat interfaces, respectively. The matters  are numbered 1, 2, and 3; the distance between the interfaces is designated as "a".   In our case it is the thickness of the oxide layer  with dielectric permittivity $\epsilon _{2}$ ; the matter 1 is the substance of the  foil with $\epsilon _{1}$;  the matter 3 is vacuum with $\epsilon _{3}=1$. The Z axis is a perpendicular to the interfaces, and the origin is placed on the interface of matters 2 and 3. For the special case when a charged particle moves along the normal to the interfaces, the spectral angular density of photons per unit solid angle after the simplification of the Pafomov's expressions is:

\begin{equation}
\label{eq1}
 \frac{ {d^2 N(\omega, \theta)}}{{d\omega d\Omega}}  = \alpha \beta^2\frac{{ sin^2\theta}}{{\pi^2 \omega }} \left| B_1 + B_2^{/}+B_2^{//}+B_3^{/}+B_3^{//}+B_3^{///} \right|^{2} \ ,
 \end{equation}

where $\alpha= 1/137.04$, $\theta$ is an emission angle  of the photons, $ B_1$ is contribution of the radiation from trajectory of the charged particle in first media taking into account multiple reflection of the radiation in media 2:

\begin{eqnarray}
\label{eq2}
 {B} _{1}  = \frac{2 {A_{||} \ 
\epsilon _{2} k_2  k_3 } }{{1 - \beta k_{1 }}} \ 
 exp\left( { - i{a\omega}/ {\beta} } \right) \ .
\end{eqnarray}

$ B_2^{/} $ and $ B_2^{//} $ are contributions of the radiation from the trajectory of the particle in the second  media in forwards and backwards direction, correspondingly  and  taking into account the multiple reflection of the radiation between first and second interfaces:

\begin{eqnarray}
\label{eq4}
 B_{2}^{/} = \frac{A_{||} ( \epsilon _{1} k_2   + \epsilon _{2} k_1) k_3 }{1 - \beta
 k_2 }{[exp{( - i a \omega k_2 ) } - exp{( - i a \omega /\beta ) } ] } \ , 
 \end{eqnarray}
 
\begin{eqnarray}
B_{2}^{//} = \frac{A_{||} ( \epsilon _{1} k_2  -\epsilon _{2} k_1)  k_3  }{1 + \beta 
k_2}{[exp{(i a \omega k_2 ) }   - exp{( - i a \omega /\beta ) } ] } \ .  
 \end{eqnarray}

$ B_3^{/} $, $ B_3^{//} $ and $B_3^{///}$ are the contributions from trajectory of the particle in the third media. $B_3^{/}$ corresponds to radiation emitted in the forward direction;  $B_3^{//}$ corresponds to the radiation emitted in the backward direction taking into account the reflecttion from the second interface; $ B_3^{///} $   corresponds  to the radiation emitted in backward direction taking into account the multiple reflection of radiation between first and second interfaces after the penetration in media 2:

 \begin{eqnarray}
\label{eq7}
 B_{3}^{/} = - \frac{{1}}{{2(1  - \beta k_3 }) } \ ,
 \end{eqnarray}

 \begin{eqnarray}
\label{eq10}
B_{3}^{//} = - \frac{  \epsilon _{2} k_3  - k_2  }{2{(1 + \beta k_3 )( k_2 + \epsilon _{2} k_3)}} \ ,  
 \end{eqnarray}

 \begin{eqnarray}
\label{eq11}
B_3^{///} = - \frac{{2A_{||} \ k_3 \epsilon _{2} k_2  }{{(\epsilon _{1} k_2   
- \epsilon _{2} k_1 ) } } }{{(1  + \beta k_3 )({{k_2 }  + \epsilon _{2} k_3  })} } exp\left( {i{a \omega k_2 } }  \right) \ .
 \end{eqnarray} 
 
 Here c = 1 and the  following notations are used:
\begin{equation}
\label{eq0}
k_{1}=\sqrt{\epsilon_1-sin^2{\theta}},\ \ k_2 = \sqrt{\epsilon_2-sin^2{\theta}},\ \  k_{3}=cos{\ \theta} \ ,
\end{equation}
\begin{eqnarray}
\label{eq12}
 A_{_{||}} = [\left( {\epsilon _{1} k_2   + \epsilon _{2} k_1 }  \right)\left( k_2   + \epsilon _{2} k_3   \right) exp\left( {-i{a \omega k_2 }} \right)\nonumber\\
 - \left( {\epsilon _{1} k_2  - \epsilon _{2} k_1  }  
\right)\left( {k_2  - 
\epsilon _{2} k_3 }  \right)  exp\left( {i{a \omega k_2 }}   \right.)]^{-1}\ .
 \end{eqnarray}
 

\acknowledgments
This work was partly supported by the program Nauka
of the Russian Ministry of Science, Grant No. FSWW
2020-0008.


\begin{thebibliography}{99}

\bibitem {aa1}
M. A. Piestrup, R. H. Pantell, H. E. Puthoff and G. B. Rothbart, \emph{Čerenkov radiation as a source of ultraviolet radiation, J. Appl. Phys.} {\bf 44} (1973) 5160.

\bibitem {aa2}
M. A. Piestrup, R. A. Powell, G. B. Rothbart, C. K. Chen and R. H. Pantell, \emph{ Čerenkov radiation as a light source for the 2000–620‐Å spectral range, Appl. Phys. Lett} {\bf 28} (1976) 92.

\bibitem {aa3}
V. A. Bazylev, V. I. Glebov, E. I. Denisov, N. K. Zhevago, A. S. Khlebnikov, V. G. Tsinoev, Yu. P. Chertov, B. I. Shramenko, \emph{Observation of Čerenkov radiation with a photon energy of 284 eV, JETP Letters} {\bf 34} (1981), 97.

\bibitem {aa4}
V. A. Bazylev, V. I. Glebov, E. I. Denisov, N. K. Zhevago, M. A.Kumakhov, A. S. Khlebnikov and V. G. Tsinoev, \emph{ X-ray Čerenkov radiation. Theory and experiment, Sov. Phys. – JETP} {\bf 54} (1981) 884.

\bibitem {aa5}
V. A. Bazylev, V. I. Glebov, E. I. Denisov, N. K. Zhevago, A. S. Khlebnikov, \emph{ Čerenkov radiation as an intense x-ray source,   JETP Lett.} {\bf 24} (1976) 371.

\bibitem {aa6}
W. Knulst, van der M. J. Wiel, O. J. Luiten and J.Verhoeven, \emph{ Observation of narrow-band Si L-edge Čerenkov radiation generated by 5 MeV electrons, Appl. Phys. Lett.} {\bf 79} (2001) 2999.

\bibitem {aa7}
W. Knulst, van der M. J. Wiel, and O. J. Luiten and J.Verhoeven, \emph{ High-brightness, narrowband, and compact soft x-ray Cherenkov sources in the water window, J., Appl. Phys. Lett.} {\bf 83} (2003) 4050.

\bibitem {aa8}
W. Knulst, J. Luiten and J. Verhoeven, \emph{ Compact, high-brightness soft x-ray Cherenkov sources, IEEE J. Sel. Top. Quantum Electron.} {\bf 10}  (2004) 1414. 

\bibitem {aa9}
W. Knulst,\emph{ PhD Thesis}, Thechnische Universit, Eindhoven (2004).

\bibitem {a10}
M. J. Moran, B. Chang, M. B. Schneider and X. K. Maruyama, \emph{ Grazing-incidence Cherenkov X-ray generation, Nucl. Instrum. Methods } {\bf B48} (1990) 287.

\bibitem {a11}
I.A. Artyukov, A.V. Vinogradov, Yu.S. Kasyanov, and S. V.Saveliev, \emph{ About X-ray microscopy in the “carbon window”, Quantum Electronics } {\bf  34}  (2004 ) 691.

\bibitem {a12}
V. A. Bazylev and  V.I. Glebov, \emph{ X-ray Cerenkov radiation at grazing incidence of electrons, Physics Letters} {\bf  A160} (6) (1991) 564.  

\bibitem {a13}
C. Gary, V. Kaplin, A. Kubankin, N. Nasonov, M. Piestrup, S. Uglov, \emph{ An investigation of the Cherenkov X-rays from relativistic electrons, Nucl. Instrum. Methods Phys. Res.} {\bf  B227} (2005) 95.

\bibitem {a14}
A. S. Konkov, P. V. Karataev, A. P. Potylitsyn and A. S. Gogolev, \emph{ X-Ray Cherenkov Radiation as a Source for Transverse Size Diagnostics of Ultra-relativistic Electron Beams, Journal of Physics: Conference Series} {\bf 517} (2014) 012003.

\bibitem {a15}
M. Shevelev, A. Konkov and A. Aryshev, \emph{ Soft-x-ray Cherenkov radiation generated by a charged particle moving near a finite-size scree Phys. Rev.} {\bf  A92} (2015) 052851.

\bibitem {a16}
M. Shevelev, A. Konkov, B. Alekseev, \emph{ Spectral and polarization characteristics of X-ray hybrid radiation, Nucl. Instrum. Methods Phys. Res.} {\bf  B464} (2020) 117.

\bibitem {a17}
M. V. Bulgakova, V. S. Malyshevsky, and G. V. Fomin,\emph{ X-Ray Cherenkov Radiation in an Absorbing Medium with Finite Dimensions Journal of  Surface Investigation. X-Ray, Synchrotron and Neutron Techniques } {\bf  14} (2020) 264.

\bibitem {a18}
 http://www.esrf.fr/computing/expg/subgroups/theory/DABAX/dabax.html

\bibitem {a19}
H. Yamada, D. Minkov, Y. Shimura, C. Scourtis, O. K. Ejike, D. Hasegawa, M. Yamada, T. Hanashima and K. Atkinson, \emph{ Measurement of angular distribution of soft X-ray radiation from thin targets in the tabletop storage ring MIRRORCLE-20SX, J. Synchrotron Rad.} {\bf  18} (2011) 702.

\bibitem {a20}
S. Uglov, A. Vukolov, V. Kaplin, L. Sukhikh and P. Karataev, \emph{ Observation of soft X-ray Cherenkov radiation in Al,  EPL (Europhysics Letters)} {\bf  118 } ( 2017) 34002.

\bibitem {a21}
V. E. Pafomov, \emph{ Radiation of a charged particle at the  interfaces presence, Trudy FIAN} {\bf  44} (1969) 28.

\bibitem {a22}
J. V. Jelley, \emph{ Cherenkov Radiation and Its Applications,} London:Pergamon, (1958).

\bibitem {a23}
L. G. Parratt, \emph{ Surface Studies of Solids by Total Reflection of X-Rays, Phys. Rev.} {\bf 95} (1954) 359.

\bibitem {a24}
A.V. Vinogradov, I.V Kozhevnikov, \emph {Multilayer x-ray mirrors, Trudy FIAN} {\bf  196} (1989) 62.

\bibitem {a25}
V.G. Kohn, \emph{ Towards the theory of specular reflection of X-rays by multilayer mirrors, Journal of  Surface Investigation. X-Ray, Synchrotron and Neutron Techniques} {\bf 1}  (2003) 23.

\bibitem {a26}
S. R. Uglov, V. V. Kaplin, A. P. Potylitsyn, L. G. Sukhikh, A. V. Vukolov, G. Kube, \emph{Investigation of the characteristics of EUV backward transition radiation generated by 5.7 MeV electrons in mono- and multilayer targets, J. Phys.: Conf. Ser.} {\bf 51} (2014) 012009.

\bibitem {a27}
http://www.esrf.eu/Instrumentation/software/data-analysis/xop2.3

\bibitem {a28}
M. R. Einbund, B. V.Polenov,  \emph{Secondary  electron multipliers and their application}, Moscow: Energoatomizdat (1981).

\bibitem {a29}
J. E. Mack, F. Paresce and S. Bowyer, \emph{ Channel electron multiplier: its quantum efficiency at soft x-ray and vacuum ultraviolet wavelengths, Appl. Optics} {\bf 15} (1976) 861.

\bibitem {a30}
A. A. Legkodymov, M. R. Mashkovtsev, A. D. Nikolenko, V. F. Pindyurin, V. V. Lyakh, S. V. Avakyan and N. A. Voronin, \emph{ Comparative of Secondary-Electron Multipliers within the Ultrasoft X-ray Range   Journal of Surface Investigation. X-Ray, Synchrotron and Neutron} {\bf 6}  (2012) 404.

\bibitem {a31}
V. I. Nefedov, \emph{ Handbook “X-ray Electron Spectroscopy of Chemical Compounds"}, Published by Chemistry, Moscow, "Khimiya" (1984).

\bibitem {a32}
E. D. Palik, \emph{ Handbook of Optical Constants of Solids III}, Academic Press, San Diego, CA (1998).

\bibitem {a33}
C.D. Wagner,W. M. Riggs, L. E. Davis, J. F. Moulder and G. E. Muilenberg, \emph{ Handbook of x-ray photoelectron spectroscopy}, Published by Perkin-Elmer (1979).

\bibitem {a34}
D. E. Cullen, J. H. Hubbell and L. Kissel, \emph{ EPDL97: The Evaluated Photon Data Library ’97 Version, Lawrence Livermore National Laboratory Report UCRL-50400} {\bf 6} (1997) 36.

\bibitem {a35}
https://physics.nist.gov/PhysRefData/FFast/Text2000/sec06.html\#tab2

\end{thebibliography}
\end{document}